\documentclass[aps,nofootinbib,preprint,showpacs,amssymb,prl]{revtex4-1}
\date{\today}
\usepackage{graphicx}
\usepackage{color}
\begin{document}

\title{Spin-aligned neutron-proton pair mode in atomic nuclei}

\author{C. Qi}
\affiliation{Department of Physics, Royal Institute of Technology,
SE-10691 Stockholm, Sweden}
\author{J. Blomqvist}
\affiliation{Department of Physics, Royal Institute of Technology,
SE-10691 Stockholm, Sweden}
\author{T. B\"ack}
\affiliation{Department of Physics, Royal Institute of Technology,
SE-10691 Stockholm, Sweden}
\author{B. Cederwall}
\affiliation{Department of Physics, Royal Institute of Technology,
SE-10691 Stockholm, Sweden}
\author{A. Johnson}
\affiliation{Department of Physics, Royal Institute of Technology,
SE-10691 Stockholm, Sweden}
\author{R.J. Liotta}
\affiliation{Department of Physics, Royal Institute of Technology,
SE-10691 Stockholm, Sweden}
\author{R. Wyss}
\affiliation{Department of Physics, Royal Institute of Technology,
SE-10691 Stockholm, Sweden}

\begin{abstract}
Shell model calculations using realistic interactions reveal that the ground and low-lying yrast states of the
$N=Z$ nucleus $^{92}_{46}$Pd are mainly built upon isoscalar neutron-proton pairs each 
carrying the maximum angular momentum $J=9$ allowed by the shell 
$0g_{9/2}$ which is dominant in this nuclear region. This structure is 
different from the ones found in the ground and low-lying yrast
states of all other even-even nuclei studied so far. 
The low-lying spectrum of excited states 
generated by such correlated neutron-proton pairs has 
two distinctive features: i) the levels are almost equidistant at low energies and ii) the transition probability
$I\rightarrow I-2$ is approximately constant and strongly selective.
This unique mode is shown to replace normal isovector pairing as the dominating coupling scheme in $N=Z$ nuclei approaching the doubly-magic nucleus $^{100}$Sn. 
\end{abstract}
\maketitle

For all known even-even nuclei it has since long been inferred that the structure of the ground states and low-lying excited states are dominated by $J=0$ isovector pairing correlations between the same kind of nucleons \cite{Boh75,Dean03}.
On the other hand, the interaction between protons
and neutrons has been found to play a key role in
inducing collective correlations and  nuclear deformation \cite{fed77,Cas06}. The isoscalar neutron-proton ($np$) interaction is dominated by a quadrupole-quadrupole component which breaks the seniority-like structure of the spectrum. As a result, many  open-shell nuclei tend to display collective
vibrations or to be deformed and exhibit rotational-like spectra while most semi-magic nuclei are spherical and dominated by the pairing (or seniority) coupling scheme. The low-lying structures of many nuclei can be understood as the outcome of the competition between these two coupling schemes (see, e.g., Ref.~\cite{Bohr80}).

Near the $N = Z$ line $np$ correlations 
may be further enhanced by the interactions arising between neutrons and
protons occupying identical orbitals. These $np$ pairs can be coupled to angular momenta $J=0$ to $2j$. The interaction energy will be strongest in the anti-parallel ($J=0$) and parallel ($J=2j$) coupling due to the large spatial overlap of wave functions. In addition, maximally aligned $J=2j$ $np$ pairs are somewhat favored by the fact that the spin-triplet nucleon-nucleon interaction is slightly stronger than the spin-singlet interaction.
In a recent work the low-lying yrast 
states in $^{92}_{46}$Pd were measured for the first time \cite{ced10}. To explain the
structure of these states it was found to be necessary to assume 
an isoscalar spin-aligned $np$ coupling scheme. It is the purpose of this Rapid Communication
to examine the role played by these $np$ pair states.
We will show that the $np$ pairs thus aligned can generate striking regular patterns in the energy 
spectra as well as in the transition probabilities along $N=Z$ nuclei. This spin-aligned $np$ pair mode with isoscalar character is unique for atomic nuclei with two distinct types of fermions. 

To illustrate the idea of the spin-aligned pair mode we will start with the
simple example of a $2n$-$2p$ system within a single $j$ 
shell. The wave function 
is usually given as the tensor product of proton and neutron pair
components, i.e.,
\begin{equation}
|\Psi_I\rangle=\sum_{J_pJ_n}X_{J_pJ_n}|[\pi^2(J_p)\nu^2(J_n)]_I\rangle,
\end{equation}
where $J_p$ and $J_n$ denote the angular momenta of the proton and neutron pairs, respectively. They can only take even values with the maximum value $J=2j-1$. The amplitude $X$ is influenced by the $np$ interaction.  One may re-express the wave function in Eq. (1) in an \textit{equivalent} representation in terms of $np$ pairs. This can be done analytically with the help of the overlap matrix as
\begin{eqnarray}\label{over}
\nonumber\langle [\nu\pi(J_1)\nu\pi(J_2)]_I |[\pi^2(J_p)\nu^2(J_n)]_I\rangle \\
= \frac{-2}{\sqrt{N_{J_1J_2}}}\hat{J_1}\hat{J_2}
\hat{J_p}\hat{J_n}\left\{
\begin{array}{ccc}
j&j&J_p\\
j&j&J_n\\
J_1&J_2&I
\end{array}
\right\},
\end{eqnarray}
where $N$ denotes the normalization factor. The overlap matrix automatically 
takes into account the Pauli principle. With interactions taken from 
experimental data and Ref. \cite{Sch76}, we examined a few shells with a high 
degeneracy, i.e., $0f_{7/2}$, $0g_{9/2}$ and $0h_{11/2}$, values in the range
$X^2_{J_p=J_n=0}=0.51-0.62$ for the ground states of these even-even nuclei. This means that 
the normal isovector pairing coupling scheme $(\nu^2)_0 \otimes (\pi^2)_0$
accounts for only about half of the ground state wave functions.
Instead, we found that for these wave functions it is 
$X^2_{J_1=J_2=2j}=0.92-0.95$, i.e., they virtually
can be represented by the spin-aligned $np$ coupling scheme.
An even more striking feature is that the low-lying yrast states
are calculated to be approximately equally spaced and their spin-aligned $np$ 
structure is the same for all of them. Moreover,
the quadrupole transitions between these states
show a strong selectivity, since the decay to 
other structures beyond the $np$ pair coupling scheme is unfavored. It should be emphasized that only states with even angular momenta can be generated from the spin-aligned $np$ coupling for systems with two pairs. The maximum spin one can get is $2(2j-1)$. For a given even spin $I$, only one state can be uniquely specified from the coupling of two aligned $np$ pairs. The other states (and also states with odd spins or total isospin $T>0$) involve the breaking of the aligned pairs.

\begin{figure}[htdp]
\includegraphics[scale=0.55]{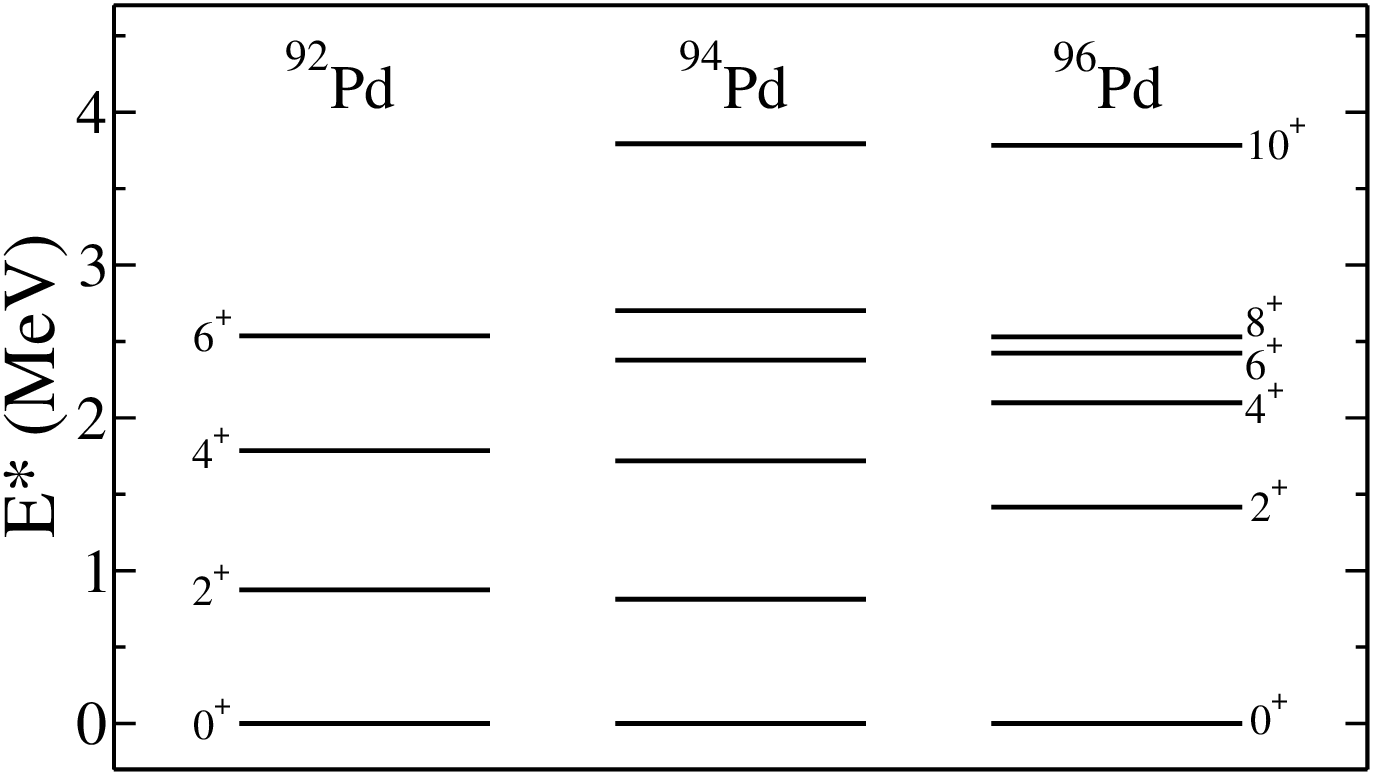}\\
\caption{
Experimental positive parity yrast states of $^{92,94,96}$Pd \cite{ced10,Nudat}.}
\label{le9094}
\end{figure}

In Fig.~\ref{le9094} 
we present the experimental spectra of three even-$A$ neutron-deficient 
Pd isotopes.
As one would expect, near the closed shell nucleus $^{100}$Sn, i.e. in
$^{96}$Pd with four proton holes with respect to  $^{100}$Sn, 
the positions of the energy levels correspond to a $(g_{9/2})_\lambda ^2$ isovector
pairing (or seniority) spectrum.  When the number of neutron holes increases, approaching $N=Z$, the levels tend to be equally separated. The lowest excited states in $^{92}$Pd exhibit a particularly regular pattern.

\begin{figure}
\includegraphics[scale=0.6]{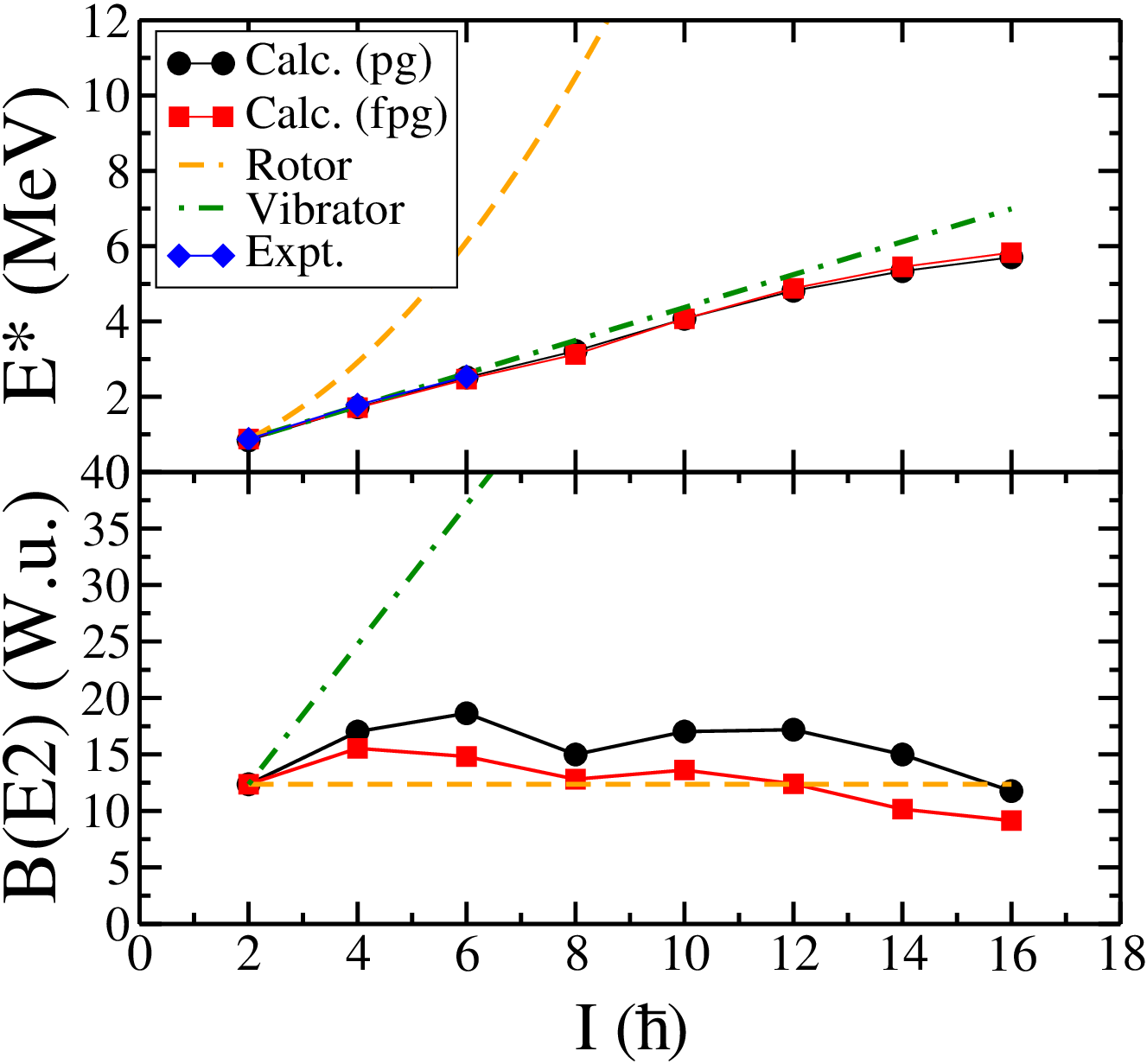}\\
\caption{(color online). 
Upper: Shell model spectra of $^{92}$Pd calculated within the 
$1p_{3/2}0f_{5/2}$$1p_{1/2}0g_{9/2}$ space \cite{hon09} ($fpg$)
and the $1p_{1/2}0g_{9/2}$ space \cite{joh01} ($pg$).
The experimental data are from Ref.~\cite{ced10}. Lower:  $B$(E2; $I\rightarrow I-2$) values in $^{92}$Pd calculated within the 
$fpg$ and $pg$ spaces. The two dashed lines show the predictions of the geometric collective model normalized to the $2^+_1$ state.}
\label{ene92pd}
\end{figure}

To understand the 
yrast structure of $^{92}$Pd, we performed nuclear shell model calculations
within the $1p_{3/2}0f_{5/2}$$1p_{1/2}0g_{9/2}$ model space ($fpg$) using 
the Hamiltonian given in Ref. \cite{hon09}.
To explore the importance of configuration
mixing in determining the structure of the spectrum we also performed
calculations using as single-particle states the orbits 
$0g_{9/2}$ and $1p_{1/2}$
($pg$) with the interaction matrix elements of Ref. \cite{joh01}. 
In the upper panel of Fig.~\ref{ene92pd} we present the calculated 
positive parity yrast states
together with the corresponding experimental data. The first outstanding feature
in this figure is the excellent agreement between theory and experiment for
all the interactions that we have used. It is important to point out that a variety of shell-model Hamiltonians have been 
constructed to explain the spectra and decay properties of nuclei in this 
mass region (for a review, see, e.g., Ref.~\cite{hon09}). Our calculations 
show that for the nuclei of concern here these Hamiltonians provide 
equivalent results. A characteristic emerging from
the calculation is that in all cases the configuration corresponding to
four maximally aligned $J=9$ isoscalar $np$ pairs is the  most important one in
the corresponding wave functions.

This can be attested by evaluating the pair content in the many-fermion 
wave function $\Psi$, that is the average number of pairs given by 
$\langle\Psi_I|((a_i^{\dag}a_j^{\dag})_{J^{\pi}}\times(a_ia_j)_{J^{\pi}})_0|\Psi_{I}\rangle$ (see, e.g., \cite{Talmi93,Qi10}). For a given coupling the average number of pairs can be larger than four and the total number of pairs is $N=n(n-1)/2=28$ in the case of $^{92}$Pd (see, Eq. (4) shown below or Ref. \cite{Qi10}). For a low-lying yrast state in $^{92}$Pd which has total isospin $T=0$, the total number of isoscalar pairs is $[n/2(n/2+1)-T(T+1)]/2=10$. If isospin symmetry is assumed, the numbers of isovector neutron-neutron, proton-proton and $np$ pairs are the same and the total number of pairs is $[3n/2(n/2-1)+T(T+1)]/2=18$. The advantage of this pair operator is that it shows in a straightforward manner the relative importance of different pairs upon the total energy \cite{Talmi93}. A more sophisticated monopole corrected pair operator also exists  (see, e.g., Ref. \cite{Cau10}) but is not employed in this work since the monopole interaction plays a relatively minor role on the wave functions and pair couplings of states discussed below.

\begin{figure}[htdp]
\includegraphics[scale=0.6]{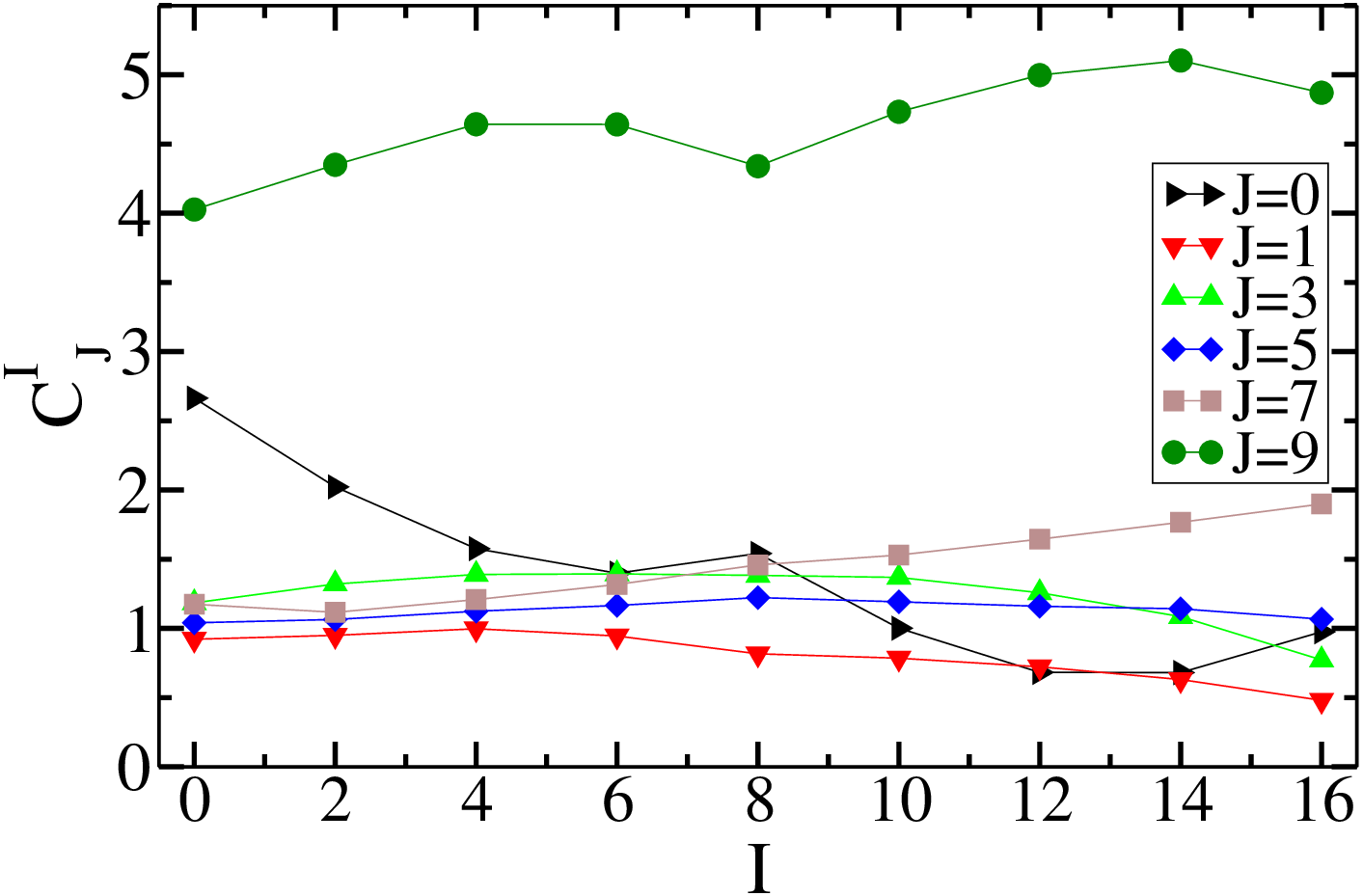}\\
\caption{(color online).
Average number of isoscalar $(0g_{9/2}^2)_J$ pairs, $C^I_J$, as a function of total angular momentum $I$ for the wave functions of the yrast states of $^{92}$Pd, in comparison with those of the normal isovector $J=0$ pair.}
\label{num}
\end{figure}

The results for the isoscalar pairs thus obtained in the $pg$ space are shown in Fig. \ref{num}. One sees that \emph{all} low-lying yrast 
states are built mainly from $J^\pi = 9^+$ spin-aligned $np$ pairs.
In contrast to the isovector pairing excitations, in the
aligned isoscalar pair state there is no pair breaking when the nucleus is
excited from one state to higher lying states in the spectrum.
Instead, the angular momentum of excited yrast states in
$^{92}$Pd is generated by the rearrangement of the angular momentum vectors 
of the aligned $np$ pairs while the isoscalar pair character itself is 
preserved. For the ground state the four pairs are maximally aligned
in opposite directions as allowed by the Pauli principle. 
When exciting the nucleus from one yrast state to
another, the dynamics of the system is determined by the active four
$(\nu\pi)_9$ hole pairs aligned differently according to the total angular
momentum of the system. It should be mentioned that the microscopic origin of the equidistant pattern generated by the aligned $np$ pairs remain unclear at present. From a simple semiclassical point of view, one may argue that the corresponding energies
vary approximately
linearly with $I$ for small total angular momenta, inducing similar energy spacings 
between consecutive levels.

In a restricted single $j$ shell, the calculations of the average pair number can be much simplified with the help of two-particle coefficients of fractional parentage (cfp) as
\begin{equation}
C^I_J=\frac{n(n-1)}{2}\sum_{\alpha\beta}X_{\alpha}M_{\alpha\beta}^I(J)X_{\beta},
\end{equation}
and
\begin{eqnarray}
\nonumber M^I_{\alpha\beta}(J) &=&\sum_{\alpha_{n-2}I_{n-2}}[j^{n-2}(\alpha_{n-2}I_{n-2})j^2(J)I|\}j^n\alpha I]\\
&&\times[j^{n-2}(a_{n-2}I_{n-2})j^2(J)I|\}j^n\beta I],
\end{eqnarray}
where $n$ is the number of particles (holes) of the system and $X$ are the expansion amplitudes of the wave function. The square bracket denotes the two-particle cfp \cite{Talmi93} which is calculated in the isospin representation in the present work. The Greek letters denote the basis states and the summation runs over all possible states. The four $J=9$ $np$ pairs in $^{92}$Pd can couple in various ways. It is thus found that the dominating components can be well represented by a single configuration $((((\nu\pi)_9\otimes(\nu\pi)_9)_{I'=16}\otimes(\nu\pi)_9)_{I''=9}\otimes(\nu\pi)_9)_{I}$. In the $0g_{9/2}$ shell, this configuration is calculated to occupy
around 66\% of the ground state wave function  of $^{92}$Pd, i.e., with amplitude $X(0^+_1)=0.81$. For the first $2^+$ state, the amplitude of above configuration is calculated to be $X(2^+_1)=0.89$.
Bearing this in mind, it may be interesting to compare present calculations with the stretch scheme proposed in Ref.~\cite{dan67}. In that
scheme the dynamics of a $2N$ system is determined by the coupling of two 
$(\nu\pi)^{N/2}$ maximally aligned (stretch) vectors. For the $4n-4p$ system of $^{92}$Pd, the corresponding  wave function for a state with total angular momentum $I$ can be written as $((\nu\pi)^2_{I_1=16}\otimes(\nu\pi)^2_{I_1=16})_{I}$ where $(\nu\pi)^2_{I_1=16}$ denotes the stretch vector. It corresponds to the unique configuration of  $((\nu\pi)_9\otimes(\nu\pi)_9)_{I_1=16}$ in the $np$ pair coupling scheme. 

\begin{figure}[htdp]
\includegraphics[scale=0.6]{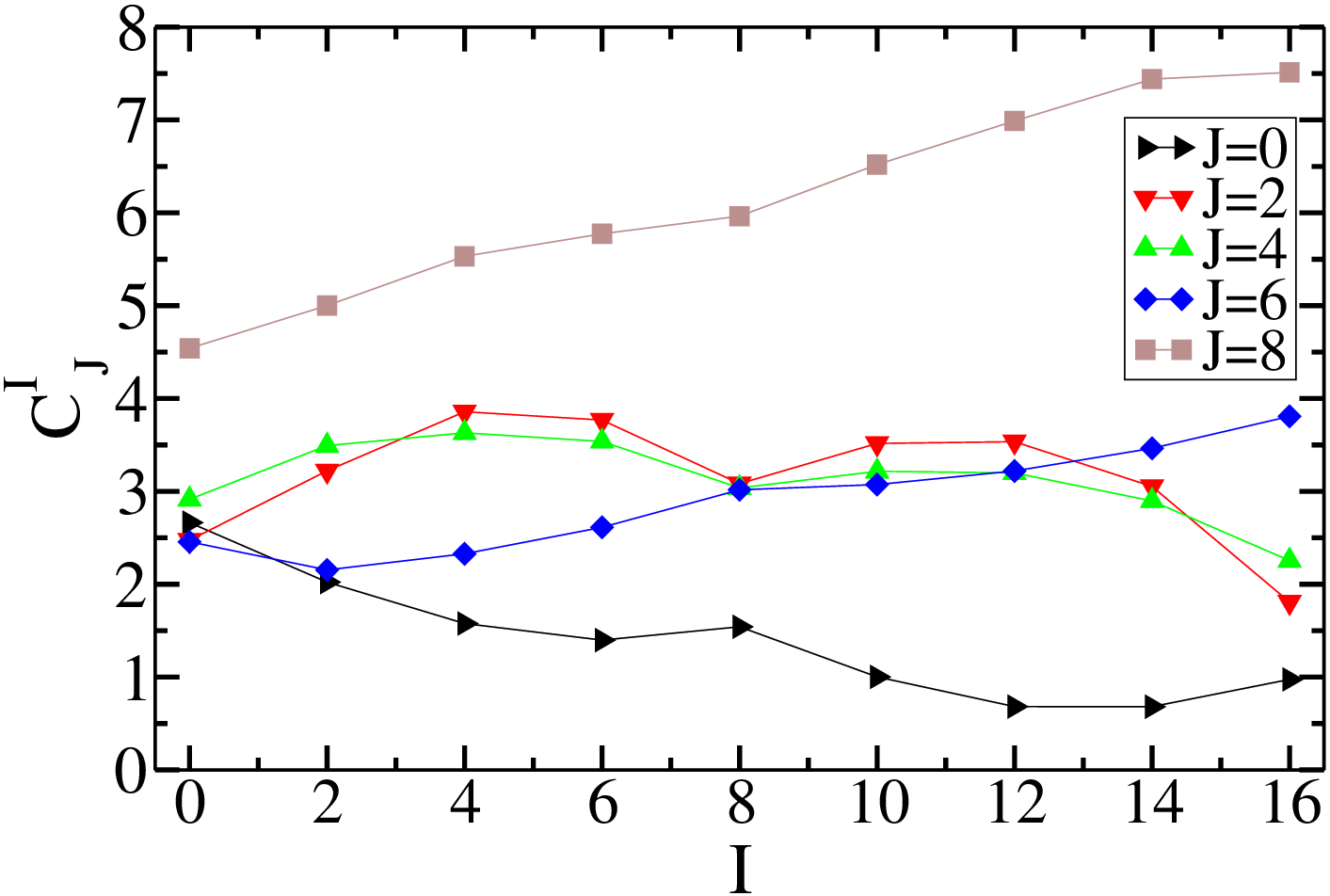}\\
\caption{(color online).
Same as Fig.~\ref{num} but for the numbers of isovector $(0g_{9/2}^2)_J$  pairs (including all proton-proton, neutron-neutron and $np$ pairs).}
\label{num2}
\end{figure}

To gain deeper physical insight into the structure of $^{92}$Pd we also calculated the numbers of isovector pairs which are plotted in Fig.~\ref{num2}. It is seen that already in the ground state we have much more pairs with $J>0$ than the normal $J=0$ pair. In particular, the dominating component is the $J=8$ pair which is maximally aligned in the isovector channel. This is due to the large overlap between states generated by the spin-aligned $np$ pairs and those generated by the isovector pairs. This phenomenon is not seen in systems with two $np$ pairs where the contributions from the isovector aligned pair is practically zero for low-lying yrast states \cite{Qi10}. It should be emphasized that the dominating component in the wave functions of low-lying yrast states in $^{92}$Pd is still the spin-aligned $np$ pair coupling. But it may be interesting to clarify the role played by the isovector aligned pair in $N=Z$ systems with more than two pairs. Work on this direction is underway.

Even though, at a first glance, the regularly-spaced level 
spectrum in $^{92}${Pd} might 
be associated with a collective vibrational motion, the calculated $B$(E2) values 
seen in the lower 
panel of Fig.~\ref{ene92pd} do not 
correspond to the predictions of the vibrational model. In fact they are
more similar to those of a rotational model. This is a peculiar feature of
the spin-aligned isoscalar coupling scheme. The sizable transition strengths between the yrast states originates from the correlated recoupling of the aligned $np$ pairs. Moreover, the multiple pair $((\nu\pi)_9)^4_I$ can be written as a 
coherent superposition (in the sense of the $B$(E2)-values)
of all isovector neutron and proton pairs occupying 
the $g_{9/2}$ shell in the 
form $((\nu^2_J)^{2} \otimes (\pi_J^2)^{2})_I$. 
In practical calculations we used for the  $fpg$ shell the standard effective charges, i.e.,
$e_p=1.5$e and $e_n=0.5$e. To facilitate comparison, the effective charges of the smaller $pg$ space are 
rescaled to reproduce the $B(E2; 2^+\rightarrow 0^+)$ value of the $fpg$ calculation.

\begin{figure}[htdp]
\includegraphics[scale=0.6]{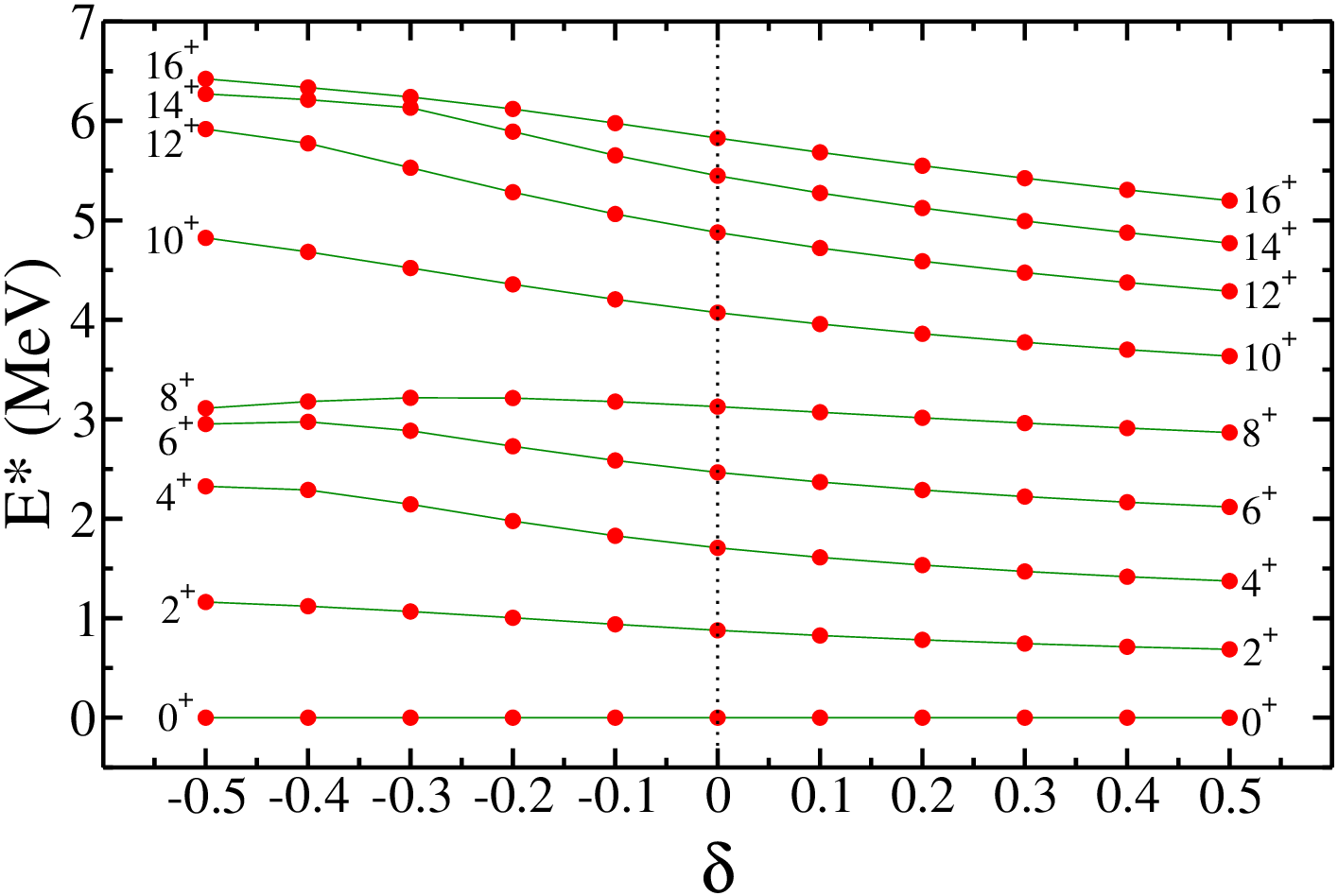}\\
\caption{(color online).
Shell model spectrum of $^{92}$Pd calculated in the 
 $1p_{1/2}0g_{9/2}$ space as a function of a controlling parameter $\delta$, i.e., we take as interaction matrix element
the value $V_9(\delta)= V_9(0)(1+\delta)$  (see text).}
\label{vdel}
\end{figure}

The main underlying mechanism behind the dominance of the maximally aligned $np$ pair mode comes from the strong attractive matrix element for the $J^{\pi}=9^+$ $np$ pair coupling, $V_9=\langle(g_{9/2}^2)_{J=9}|V|(g_{9/2}^2)_{J=9}\rangle\sim-2$~MeV, 
because of the large spatial overlap of the particle distributions. 
This is illustrated in Fig. \ref{vdel} where the energy
levels of $^{92}$Pd are calculated from a fractional change of $V_9$ by a
quantity $\delta$, that is by taking as interaction matrix element
the value $V_9(\delta)= V_9(0)(1+\delta)$ where $V_9(0)$ denotes the one given in the original Hamiltonian.
The figure reveals that an increasing $V_9(\delta)$ results in
low-lying yrast states which are approximately equidistant. 
In contrast, as $V_9(\delta)\rightarrow 0$ a seniority-like situation 
is reached. The pattern is analogous to that of Fig.~\ref{le9094}. This 
coincides with the analysis of the $np$ interaction \cite{ced10}, but now we see that it is
indeed the pair mode $(g_{9/2}^2)_9$ that dominates the spectrum. 
 For simplicity, the calculations in Fig.~\ref{vdel}  refers to the $pg$ space~\cite{joh01}. The minor effect of this variation on the $T=0$ monopole interaction is not taken into account. Calculations in the restricted $0g_{9/2}$ single shell and extended $fpg$ space show a similar pattern.

We have found  a similar coupling scheme and excitation pattern for the calculated low-lying yrast states 
of $^{96}$Cd. The calculated yrast spectrum exhibits a equally spaced pattern up to spin $I=6$. Here the lower spin 
states (and also the highest spin states) are completely dominated by the aligned $np$ paired scheme  
$|((\nu\pi)_9\otimes(\nu\pi)_9)_I\rangle$. The situation becomes more complicated in states with spin around $I=8$ due to the competing effect from the isovector aligned pair $(0g_{9/2})^2_8$ (which is related to the kink in the calculated number of pairs along the yrast line \cite{Qi10}). In particular, the calculated $8^+_1$ state favors the non-collective seniority-like $((0g_{9/2})^2_0\otimes(0g_{9/2})^2_8)_{I=8}$ configuration rather than the aligned $np$ pair coupling. As a result, the $E2$ transition from this state to the $6^+_1$ state is hindered. Our calculations show that it is the third $8^{+}$ state (the second $T=0$, $8^+$ state) that corresponds to the spin-aligned $np$ pair coupling. The calculated spectrum of $^{96}$Cd follows, as a function of 
the controlling parameter $\delta$, a pattern similar to that of 
Fig.~\ref{vdel}. By enhancing the $(0g_{9/2})^2_9$ pair interaction a transition from the seniority coupling to the aligned $np$ pair coupling is seen in the $8^+_1$ state and  the equally-spaced pattern is extended to higher spins.

The proposed spin-aligned $np$ coupling may also shed some new light on the structure of odd-odd $N=Z$ nuclei. Here we take the nucleus $^{94}$Ag as an example. It has three $np$ hole pairs in the $^{100}$Sn core. The ground state is calculated to be $J^{\pi}=0^+$ and $T=1$ and is dominated by the coupling of two aligned $np$ pair and a normal isovector $J=0$ pair. In contrast to the two-pair case, systems with three aligned $np$ pairs can coupled to both even and odd spin values. For examples, shell-model calculations \cite{joh01,hon09} give two low-lying $T=0$ states. One is a $7^+$ state with excitation energy $E*\sim0.7$ MeV, and the other one is a $8^+$ state ($E*\sim0.9$ MeV). Both states are dominated by the aligned $np$ pair coupling of $(((\nu\pi)_9\otimes(\nu\pi)_9)_{I'=16}\otimes(\nu\pi)_9)_{I}$ with $|X|\approx 0.97$ in restricted calculations performed within the $0g_{9/2}$ shell. This component decreases to $|X|=0.83$ for the first $9^{+}$ state, which is otherwise dominated by the coupling $(((\nu\pi)_9\otimes(\nu\pi)_9)_{I'=0}\otimes(\nu\pi)_9)_{I}$. Again the wave functions of these states are mixtures of many components if we express them in the normal proton-proton and neutron-neutron coupling scheme.


In summary, we have shown that the recently measured spectrum 
of the $N=Z$ nucleus $^{92}$Pd, with four neutron and four proton holes in the 
$^{100}$Sn core, can best be explained in terms of maximally aligned isoscalar $np$
pairs. We predict also that such structures dominate the ground and
low-lying yrast states of the heavier $N=Z$ nucleus $^{96}$Cd.
To reach this conclusion we first noticed that in
these nuclei the
single-particle shell $0g_{9/2}$ determines the spectra. Moreover, the most
important quantity controlling the spectrum is the matrix element that corresponds to the maximal angular momentum
alignment resulting in $np$ pairs with $J=9$.
The strong attractive aligned $np$ pair mode 
overwhelms the normal isovector pairing as the dominating coupling scheme in 
the ground states of $^{92}$Pd and $^{96}$Cd.
The near equidistant spectrum measured in $^{92}$Pd is due
to the relative motion of spin-aligned $np$ pairs (each of which acts as
a frozen degree of freedom). 
This mode of excitation is unique in nuclei and indicates that the $np$
spin aligned pair has to be considered as an 
essential building block in nuclear structure calculations. Measurements of E2 transition rates are needed to really confirm this interpretation.

This work has been supported by the Swedish Research Council (VR) under grant 
Nos. 623-2009-7340, 621-2010-3694 and 2010-4723.

\end{document}